\documentclass[iop]{emulateapj}

\shorttitle{Chemistry of dwarfs in NGC~2808}
\shortauthors{Bragaglia et al.}

\begin{document}
\title{X-shooter observations of main sequence stars in the globular cluster NGC~2808: first chemical tagging of a He-normal and a He-rich dwarf
\altaffilmark{1}}

\author{A. Bragaglia\altaffilmark{2}, E. Carretta\altaffilmark{2},
R.G. Gratton\altaffilmark{3},
S. Lucatello\altaffilmark{3,4,5},
A. Milone\altaffilmark{6},
G. Piotto\altaffilmark{6},
V. D'Orazi\altaffilmark{3},
S. Cassisi\altaffilmark{7},
C. Sneden\altaffilmark{8},
L.R. Bedin\altaffilmark{9}}

\altaffiltext{1}{Based on data collected at the ESO telescopes under  GTO programme 084.D-0070}
\altaffiltext{2}{INAF, Osservatorio Astronomico di Bologna, via Ranzani 1,
       40127,  Bologna,  Italy. angela.bragaglia@oabo.inaf.it, 
       eugenio.carretta@oabo.inaf.it}
\altaffiltext{3}{INAF, Osservatorio Astronomico di Padova, vicolo
       dell'Osservatorio 5, 35122 Padova,  Italy. raffaele.gratton@oapd.inaf.it
       sara.lucatello@oapd.inaf.it, valentina.dorazi@oapd.inaf.it}
\altaffiltext{4}{Excellence Cluster Universe, Technische Universit\"at M\"unchen, 
Boltzmannstr. 2, D-85748, Garching, Germany}
\altaffiltext{5}{Max-Planck-Institut fuer Astrophysik, D-85741 Garching, Germany}
\altaffiltext{6}{Dipartimento di Astronomia, Universit\`a di Padova, Vicolo dell'Osservatorio 2,
I-35122 Padova, Italy. antonino.milone@unipd.it, giampaolo.piotto@unipd.it}
\altaffiltext{7}{INAF, Osservatorio Astronomico di Collurania, via M.Maggini, 
I-64100 Teramo, Italy. cassisi@oa-teramo.inaf.it}
\altaffiltext{8}{Dept. of Astronomy and McDonald Observatory, The University of
Texas, Austin, TX 78712, USA. chris@verdi.as.utexas.edu}
\altaffiltext{9}{Space Telescope
Science Institute, 2700 San Martin Drive, Baltimore, MD 21218, USA}

\begin{abstract}
We present the first chemical composition study of two unevolved stars in 
the globular cluster NGC~2808, obtained with the spectrograph X-shooter@VLT. NGC~2808 
shows three discrete, well separated main sequences. The most accepted explanation 
for this phenomenon is that their stars have different helium contents. We observed
one star on the bluest main sequence, (bMS, claimed to have high helium content,
Y$\sim 0.4$), and one on the reddest main sequence (rMS, consistent with a canonical
helium content, Y=0.245). We analyzed features of NH, CH, Na, Mg, Al, and Fe. While Fe, 
Ca, and other elements have the same abundances in the two stars, the bMS star 
shows a huge enhancement of N, a depletion of C, an enhancement of Na and Al, and
small depletion of Mg with respect to the rMS star. This is exactly what 
is expected if stars on the bMS formed from the ejecta produced by an earlier stellar
generation in the complete CNO and MgAl cycles whose main product is helium. The elemental
abundance pattern differences in these two stars are consistent with the differences
in helium content suggested by the color-magnitude diagram positions of the stars.
\end{abstract}


\section{Introduction}

The main sequence (MS) of the globular cluster (GC) NGC~2808 shows a wide color
distribution (\citealt{dantona05}). Very accurate HST ACS photometry 
(\citealt{piotto07}) reveals that the MS splits into three sequences that are 
likely formed in discrete episodes of star formation, slightly separated in age, 
and with different initial chemical composition. The MS of the massive,
multi-metallicity GC $\omega$~Cen also is split in two \citep{bedin04} or even 
more \citep{bellini10} separate branches. NGC~2808 and $\omega$~Cen are
presently the only GCs where the MS is clearly separated into discrete
sequences; see the review by \cite{piotto09}.

The main r\^ole in producing these broad MSs is probably played by helium
\citep[e.g.,][]{norris04}. From 
the analysis of stacked spectra of 17 stars  in each sequence of $\omega$~Cen, \cite{piotto05} 
clearly showed that the blue main sequence (bMS) is more metal-rich than the red 
MS (rMS). The most reasonable way to reconcile these observations with stellar 
evolutionary theory is to suppose that the bMS is populated by stars born with 
larger helium abundance than the rMS stars. By analogy, He variations provide
the simplest explanation for the three distinct NGC~2808 MSs, since the separation 
cannot be explained easily by age and/or metallicity variations. The three MSs 
can be fit by theoretical models with He content ranging from a {\it normal} Y=0.24 to 
an extreme Y=0.38 value (see Fig. 2 in \citealt{piotto07}).

Helium abundances cannot be directly measured by spectroscopy, except for high 
temperature, highly evolved stars \citep[e.g.,][]{moehler06,villanova09}.
Fortunately, spectroscopic investigations of GCs over several decades have
found large star-to-star and cluster-to-cluster variations in other light
element abundances, which strongly indicate the existence of multi-populations.
In all clusters studied so far\footnote{Possible exceptions are the poorly-studied 
GCs Ter~7 and Pal~12 \citep{sbordone,cohen}, where however only seven and four 
stars were studied, respectively.} large star-to-star variations in light elements 
O, Na, Mg, Al, Si are present (e.g., \citealt{carretta09a,carretta09b}, and see 
\citealt{gratton04} for a recent review).  The variations have distinctive patterns:
O and Mg abundances are
positively correlated, and are anti-correlated with Na, Al, and Si
abundances.  Such patterns leave little doubt about the chief
nucleosynthesis culprit: high-temperature hydrogen fusion that includes CNO,
NeNa, and MgAl cycles. Moreover, Na-O and Mg-Al anticorrelations have been found
among {\it unevolved} MS and subgiant stars (\citealt{gratton01}), arguing that
this pattern is produced by the ejecta of a first generation of now extinct more 
massive stars (\citealt{den89,langer93}). The site of the H burning is still unclear.
It might have occurred either in intermediate-mass asymptotic giant branch (AGB) 
stars (\citealt{dantona02}) or in fast rotating massive stars on the MS
(\citealt{decressin07}).

The main outcome of H burning, helium, is expected to be directly related to the 
observed chemical pattern of light elements in GCs. Stars on the MS that are
highly enriched in He should have large depletions of O and Mg, and large
enhancements of Na and Al (possibly Si as well). They also should populate
the extreme blue horizontal branch (HB) (see \citealt{bragaglia10} and 
\citealt{gratton10}) and the bluer MS. 

NGC~2808 is the ideal target to derive a clearcut confirmation of these effects.
The close link between He-enhancement and the simultaneous depletion/increase 
in light elements may be investigated in this GC because: (i) the three {\it 
distinct} MSs allow unambiguous selection of stars with likely different He 
content; (ii) this cluster has an unusual HB, strongly bimodal (or even trimodal), 
and with a long blue (hot) tail (e.g. \citealt{bedin00}) that has been connected 
to different He enrichments (\citealt{dantona04}); (iii) it shows a very extended 
Na-O anticorrelation (\citealt{carretta06}), also implying  very large 
He-enhancements (see also \citealt{bragagliahe}). As discussed by \cite{piotto07}, 
there seems to be a clear connection between the various groups of MS, RGB, and 
HB stars, according to star counts in the studies of \cite{piotto07}, 
\cite{carretta06}, and \cite{bedin00}, respectively. 

With the advent of X-shooter \citep{vernet09} at VLT, medium resolution spectra useful 
for abundance analysis are currently within reach for faint MS stars of NGC~2808. 
In this Letter we present the first results of the chemical tagging of two stars 
observed {\it in situ} on the bluest and the red MS of this cluster.

\section{Observation, reduction, and analysis}

\begin{table}
\centering
\setlength{\tabcolsep}{1.2mm}
\caption{Information on the two targets and derived abundances.}
\begin{tabular}{llcc}
\tableline
Star        &       & rMS$-$star & bMS$-$star \\
\tableline
RA          &(h:m:s)& 09:11:36.29 &09:11:30.80 \\
Dec         &(d:p:s)&$-$64:55:19.17 &$-$64:53:22.49 \\
$m_{\rm F475W}$       &(mag)  & 20.465      &20.416 \\
$m_{\rm F814W}$       &(mag)  & 19.232      &19.230 \\
T$_{\rm eff}$&(K)   & 6252        &6479 \\
$\log{g}$   &       & 4.32        &4.27 \\
$[$C/Fe$]$      &       & $-$0.3 &$-$0.7 \\
$[$N/Fe$]$     &       & +0.5 &+2.0 \\
$[$Na/Fe$]$     &       & $-$0.3 &+0.8 \\
$[$Mg/Fe$]$     &       & +0.4 &+0.1 \\
$[$Al/Fe$]$     &       & $-$0.2 &+1.1 \\
\tableline
\end{tabular}
\tablecomments{For both stars we used [Fe/H]=$-1.1$ and $v_{\rm t}$=0.8 km~s$^{-1}$. 
The adopted distance modulus and reddening are $(m-M)_V=15.0$, $E(B-V)=0.18$. }
\label{info} 
\end{table}

\begin{figure} 
\includegraphics[width=9cm]{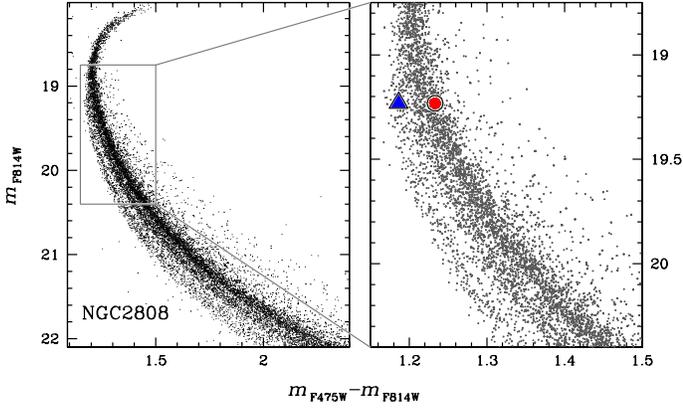}
\caption{The triple main sequence of NGC 2808 observed with ACS@HST,
corrected for differential reddening \citep{piotto07}.  Our targets are indicated with a large circle
(rMS-star) and a large triangle (bMS-star) in the enlargement (right panel).}
\label{cmd}
\end{figure}

We selected our targets among the brightest stars for which the separation of the
MSs is still possible, since the three MSs merge near the turn-off of the cluster 
(Fig.~\ref{cmd}). As confirmed on theoretical grounds \citep{scbook}, the combined 
effects of larger brightness and shorter lifetime during the MS stage when the 
initial He content increases, largely cancel out. As a consequence two isochrones 
of same age and different He contents almost overlap in the turnoff and subgiant 
regions. From the ACS photometric catalogue \citep{piotto07} we selected two MS 
stars in NGC~2808 that: i) have proper motions typical of cluster stars, ii) have 
a probability $>85\%$ of being on the bMS and $>95\%$ of being on the rMS, 
based on their magnitude and colors, respectively, and iii) do not have 
neighbours closer than 1.5\arcsec. 

As part of the Italian X-shooter guaranteed time observations (GTO), we observed 
these two MS stars with the X-shooter spectrograph at VLT-UT2 on January 22-23 2010.
The wavelength coverage of X-shooter ranges from the atmospheric UV limit to the 
near infrared. For our observations, the slit width was set at 1\arcsec or 0.8\arcsec, 
but the resolution ($R\sim 10000$) was mostly dictated by the sub-arcsec seeing, 
especially for the second night. We took four and five 1-hour exposures for star 5 
and 6, respectively. Information on the targets is given in Table~\ref{info} and 
their position in the color-magnitude diagram (CMD) is shown in Fig~\ref{cmd}. 
The observations were optimized for the UV and visual part of the spectrum, i.e., 
we did not use the nod option, essential to subtract the sky in IR. Therefore, we 
use here mostly the UV/blue spectrum ($\sim$3300-5500~\AA), and part of the visible 
($\sim$5500-10000~\AA). The spectra were reduced using the preliminary X-shooter 
pipeline (v0.9.4) (see \citealt{goldoni06}) and standard IRAF\footnote{IRAF is  distributed by the National 
Optical Astronomical Observatory, which are operated by the Association of 
Universities for Research in Astronomy, under contract with the National Science 
Foundation} routines. Each stellar exposure was bias-corrected and flat-fielded, calibrated in wavelength, then extracted and corrected for sky background. Our targets 
are near the limit of the instrument response, and the sky level is comparable to 
the star signal (or higher than it, in presence of even a small Moon contribution).
The  individual spectra have S/N from about 10  to about 20. They were combined, 
weighting them with their S/N, and shifted to zero radial velocity. The radial
velocities were measured using about 35 lines; the heliocentric values are 75 and 
80~km~s$^{-1}$ (r.m.s. 10~km~s$^{-1}$), for star rMS-star and bMS-star, respectively. Given 
the resolution and the uncertainty due to the use of a (not filled) slit, they well 
compare to the value of 93.6~km~s$^{-1}$ reported by \cite{harris96}.

To derive temperatures and gravities we used a combination of photometric
information  and isochrone fitting. Isochrones \citep{pietrinferni06}, distance
modulus, and  reddening are those used in \cite{piotto07} to fit the discrete
MSs observed in NGC~2808. The isochrones consistent with the rMS position of the
first star has a canonical  He abundance (Y=0.248), and that for the bMS
position of the second star has enhanced He (Y$\simeq$0.40).  In the current
interpretation rMS-star is of first-generation and bMS-star of
second-generation. The individual values for T$_{\rm eff}$ and $\log{g}$  are
indicated in  Table~\ref{info}. For both stars we adopted the NGC~2808
metallicity derived  from high-resolution spectra by \cite{carretta06}, [Fe/H]=$-1.1$, and assumed a
microturbulent velocity v$_t$=0.8 km s$^{-1}$. These are reasonable assumptions,
given the identical evolutionary states and the  internal homogeneity in heavy
elements and metallicity\footnote{With very few exceptions,  the stars
metallicity is very homogeneous in GCs (better than 10\%,
\citealt{carretta09c}).}.  The exact value of v$_t$ does not strongly influence
the results presented here.

The main interest of our analysis lies in the differential analysis 
of the two stars.
However, the temperatures adopted are supported by the Balmer lines, and 
we confirmed the assumed metallicity by computing synthetic spectra near 
a few Fe {\sc i} lines with good $gf$ values. We did not
explicitly took into account an enhancement in He in the stellar atmospheres;
however,
the expected effect on iron abundance determination is negligible
\citep[see discussions in][]{carretta06,bragagliahe} and this is most probably
true for the other elements.
 
Although we did not do a detailed
analysis of other heavy-element species, we noted that lines of e.g., Ti and Ca
have approximately the same strength in both stars. None of the suggested pollution
sources that contribute hydrogen-burning products to newly forming stars should
contribute elements beyond Si. Heavier $\alpha$\ or Fe-peak abundances should
be the same in all NGC~2808 stars, and our spectra do not contradict this
expectation.

\section{Results}

Adopting the stellar parameters defined above, we used standard routines to
compute  synthetic spectra for some particularly interesting elements:  N (from
the NH feature  at 3360~\AA), Al (from the 3961~\AA\ resonant line), Mg (from
the Mg b lines near  5180~\AA), C (from the CH features in the G-band near
4300~\AA), and Na (from the  8183-94~\AA\ doublet). The spectral syntheses shown
in the figures were computed  with the LTE spectroscopic analysis code ROSA
\citep{rosa} for the atomic lines (Al,  Mg, and Na) and MOOG \citep{moog} for
the molecules (NH, CH). However, all the synthesis  work was checked
independently using both codes; results are in very good agreement.

Abundances for all elements are presented in Table~\ref{info}. The derived
values have conservative error estimates (mostly due to the uncertain continuum
placement)  of 0.1~dex for Na (and Fe), and 0.2~dex for N, C, Mg, and Al. We
stress however that the main  result of our analysis lies in the {\em
difference} between the abundance patterns  of the two stars, more than in the
absolute values for the chemical abundances. We will show that the light-element
differences between rMS-star and bMS-star exceed their uncertainties.

\begin{figure} 
\includegraphics[width=15cm]{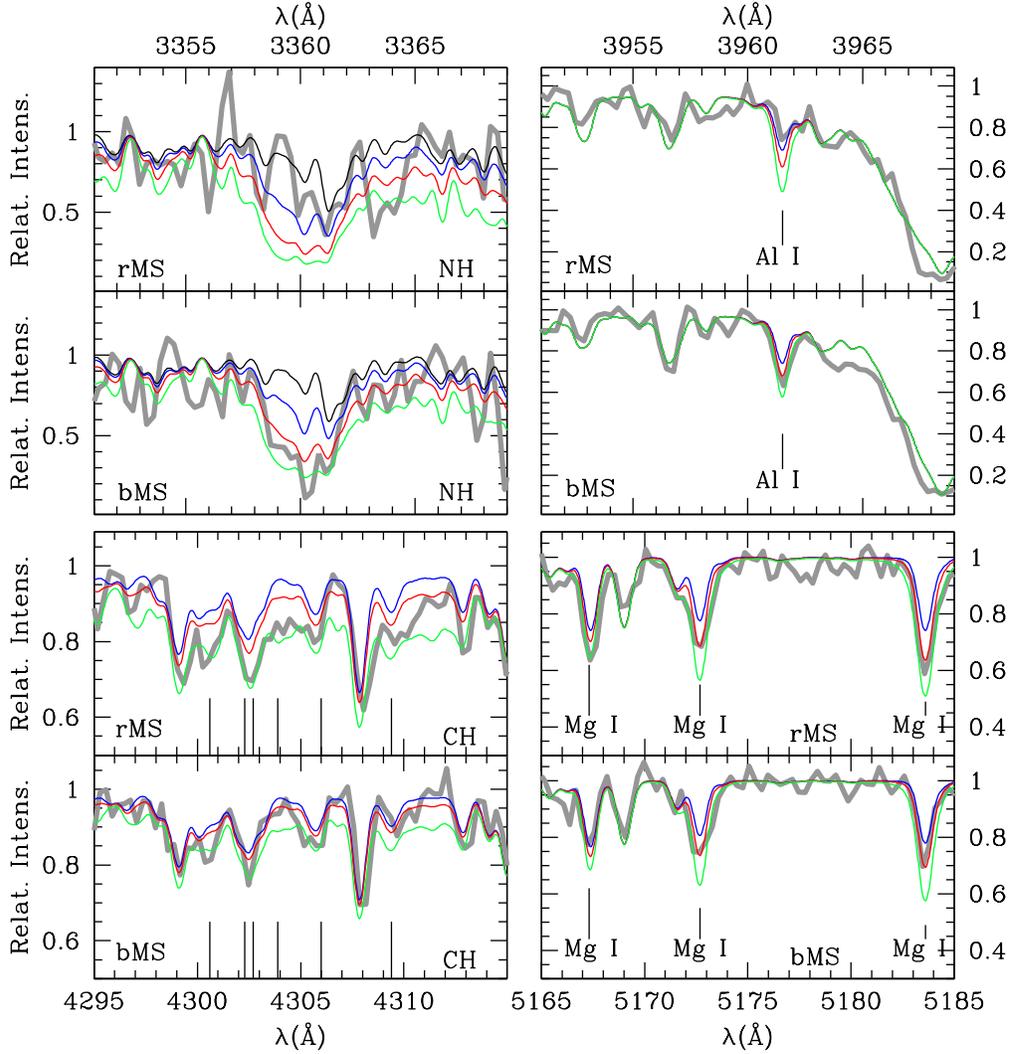}
\caption{Synthetic (light lines) and observed spectra (heavier, grey lines)
for rMS-star  (upper panels) and bMS-star (lower panels) for NH, Al, Mg,
and CH, clock-wise from the upper left panel. All spectra were normalized to
unity. In all panels the vertical lines indicate the spectral lines synthesized. The
different synthetic spectra were computed with the following abundances: a)
[N/Fe]= 0, 1.0, 1.5, 2.0; [Al/Fe]= $-0.7$, $-0.2$, 0.3, 0.8 (LTE); 
[Mg/Fe]= $-0.5$, 0.0, 0.5; 
[C/Fe]=$-1.0$, $-0.5$, 0.0.}
\label{synth}
\end{figure}

Comparison of observed and synthetic spectra are shown in Fig.~\ref{synth} for
NH,  Al {\sc i}, CH, and Mg {\sc i} features. From the closest
observed/synthetic matches  we estimated the abundances that are given in
Table~\ref{info}.   The expectations are that N and Al should be increased, and
C and Mg should be decreased in bMS-star with respect to the values for rMS-star
(the one of supposedly {\it normal}, primordial composition), following what has
been found for evolved RGB stars  (e.g.,
\citealt{ivans01,cohen02,rc03,carretta09b}). Fig.~\ref{synth} demonstrates that 
the two stars have different light-element spectra. This is most obvious for N:
the NH  absorption is much stronger in bMS-star than in rMS-star. Neither
details in the spectrum  normalization nor (small) differences in the
atmospheric parameters can account for this difference. 

The Al abundance was derived only from the 3961~\AA\ resonance line, since its
doublet partner at 3944~\AA\ is a blend \citep{arpigny83}. The synthesis was
computed  adopting the Ca abundance appropriate for NGC~2808 ([Ca/Fe]=+0.34,
\citealt{carretta09b})  to reproduce the Ca {\sc ii} H \& K lines. The values
for Al and Mg  given in  Table~\ref{info} are corrected for NLTE effects
according to \cite{gehren04}; the  corrections are about +0.5 dex for Al and
+0.06 dex for Mg, respectively, for both stars.

The huge abundance of N found for bMS-star (which also had decreased C) can be
explained only with the transformation of (virtually) all oxygen into nitrogen.
Our findings seem to indicate that we are seeing, in the gas from which this
star formed, the outcome of the complete CNO cycle. Actually, if we combine the
[C/Fe] and [N/Fe] values of Table~\ref{info} with the solar C, N, and O
abundances by \cite{asplund09} and with the maximum [O/Fe] ratio for RGB stars
in NGC~2808 (\citealt{carretta06}), even large depletions of O ([O/Fe]$<-1$)
cannot reproduce a constant sum of the CNO elements. This would be reproduced by
assuming [N/Fe]$\sim 1.4$ for bMS-star. We note that a systematic offset of
$\sim 0.6$~dex in N abundances from our analysis would produce  a roughly solar
scaled [N/Fe] ratio for rMS-star, which would agree fairly well with the values
usually assumed for field halo stars (\citealt{gratton00}). 
Once again, what is most important is the difference between the derived
abundances, and this is a sound result.

\begin{figure} 
\includegraphics[bb=40 170 310 430, clip, width=8cm]{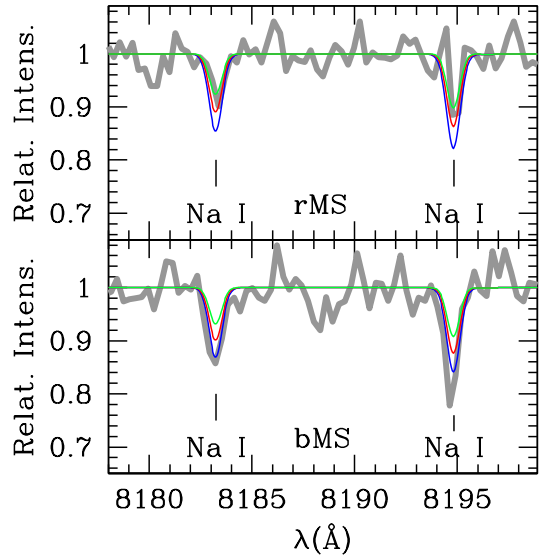}
\caption{Observed (grey, thick lines) and synthetic (thinner lines) spectra
for the Na features at 8183-94 \AA.  The synthetic spectra are for 
[Na/Fe]=0.0, 0.4. 0.8 (LTE).}
\label{na}
\end{figure}

Unfortunately, it was not possible to measure O abundances for these two stars, since 
the O triplet at 7771-7774~\AA\ is weak and falls in a wavelength region where the sky subtraction 
is difficult for these very faint objects. However, we were able to estimate the Na abundances. 
Fig.~\ref{na} shows observed and synthetic spectra surrounding the 8183-94~\AA\ Na {\sc i} lines. 
We see in the figure that the Na lines are stronger in bMS-star than in rMS-star and this 
is reflected in the abundance ratios indicated in Table~\ref{info}. These abundances include
NLTE corrections (about -0.1~dex) as reccomended by \cite{gratton99}. The chief
limitation is the strong telluric-line contamination in this spectral region. The
tellurics were eliminated by division of the program star spectra with that of a 
hot, rapidly rotating star, using an IRAF routine for this task. The cleaning quality 
is much better for the bluest of the two lines, so that our Na abundances rest on that 
single feature. They are qualitatively confirmed by the other line and by the relative strengths of the Na~D lines.
Unfortunately, strong interstellar absorption and sky emission made the derivation of Na 
abundance from the D lines less secure, given the radial velocity at the time of 
observation and the moderate resolution of the X-shooter spectra. 

\begin{figure} 
\includegraphics[width=8cm]{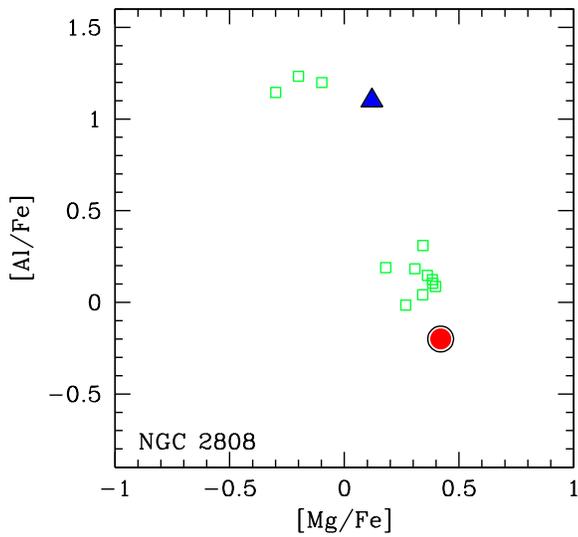}
\caption{Anticorrelation between Al and Mg abundances. The
open squares are for 12 RGB stars analysed in \cite{carretta09b}, while the
filled circle and triangle indicate the values for rMS-star and bMS-star,
respectively. }
\label{almg}
\end{figure}

The present Al and Mg results are in good agreement with those found from high resolution UVES 
spectra of 12 red giants in NGC~2808 by \cite{carretta09b}. In Fig.~\ref{almg} we plot the 
Mg-Al anticorrelation for the RGB stars and the two MS stars analyzed here. 
Apart from a possible small zero-point effect due to the use of different lines,
corrections for NLTE, etc., the two MS stars do nicely participate in the same 
trend defined by the giants.
The rMS and the bMS star fall in the Mg-rich/Al-poor and Mg-poor/Al-rich groups, respectively.
This result indicates that the extreme abundance pattern of the Al-rich, Mg-poor
stars in NGC~2808 must have been produced by pollution from a previous stellar generation in 
the clusters. Deep mixing has been recently revisited as an explanation for the extreme chemical 
abundances of RGB stars in GCs (see, e.g., \citealt{dv07,lee10}). However, since deep
mixing cannot have been 
responsible for the light element abundances in the bMS star, it is unlikely to have caused 
the identical pattern in evolved RGB stars of NGC~2808.

Finally, it is interesting to note that the Ba abundance, as indicated by the
resonance Ba line at 4554~\AA, seems the same for the two MS
stars. If confirmed by more quantitative analysis, this would probably
exclude a significant contribution from low-mass AGB stars \citep[e.g.,][]{yong09} to the pool of gas
from which the second-generation stars originated. The consequence is that the
handful of stars observed to have strong enhancement in Ba and other $s-$process
elements in some GCs must have another origin, likely 
from mass transfer from a former AGB companion in a binary system (see \citealt{dorazi10}).

\section{Conclusions}

In conclusion, the chemical pattern we found from the first abundance analysis 
of a star on the He-rich MS sequence of NGC~2808 is exactly what is expected if
stars on the bMS formed from ejecta produced by an early stellar generation via
proton-capture reactions in H-burning at high temperature, accompanied the
main outcome of this nuclear burning, i.e. helium. The extreme Al enhancement
and Mg depletion observed in the bMS star argues against a deep mixing hypothesis
for the extreme chemical abundances of RGB stars in GCs.

Observations of a larger sample of unevolved stars in this cluster and others with
suspected He variations would be welcome.

\acknowledgements

We thank Paolo Molaro for the nice observations made on behalf of the Italian 
X-shooter GTO team and Valentina D'Odorico for help and suggestions with the 
data reduction. The organizing work of Paul Groot and Sofia Randich is acknowledged. 
Funding come from PRIN-MIUR 2007, the Italian GTO X-shooter Consortium, and US 
National Science Foundation grant AST-0908978. This paper is dedicated to Roberto 
Pallavicini, late Italian PI of the X-shooter Consortium, who spent a lot of time 
and work for the realization of this fine instrument.

\clearpage

\begin{thebibliography}{}

\bibitem[Arpigny \& Magain (1983)]{arpigny83}
Arpigny, C., \& Magain, P., 1983, \aap, 127, L7

\bibitem[Asplund et al. (2009)]{asplund09} Asplund, M., Grevesse, N., Sauval, A.J., Scott, P., 2009, ARA\&A, 47, 481

\bibitem[Bedin et al. (2000)]{bedin00} 
Bedin, L., Piotto, G., Zoccali, M.,
Stetson, P.B., Saviane, I., Cassisi, S., Bono, G. 2000, \aap, 363, 159

\bibitem[Bedin et al.(2004)]{bedin04} 
Bedin, L.~R., Piotto, G.,  Anderson, J., Cassisi, S., King, I.~R., Momany, Y., \& 
Carraro, G.\  2004, \apjl, 605, L125 

\bibitem[Bellini et al.(2010)]{bellini10} 
Bellini, A., Bedin, L.R., Piotto, G., Milone, A.P., Marino, A.F., Villanova, S.
2010, AJ, in press, arXiv:1006.4157

\bibitem[Bragaglia (2010)]{bragaglia10}
Bragaglia, A. 2010, in IAU Symp. 268 ``Light elements in the Universe", ed. C.
Charbonnel, M. Tosi, F. Primas, \& C. Chiappini (Cambridge: Cambridge Univ.
Press), 119

\bibitem[Bragaglia et al.(2010)]{bragagliahe}
Bragaglia, A., Carretta, E., Gratton, R., D'Orazi, V., Cassisi, S., \& Lucatello, S. 2010, \aap, 
in press, arXiv:1005.2659

\bibitem[Carretta et al. (2006)]{carretta06} Carretta, E., Bragaglia, A., Gratton R.G., Leone, F., 
Recio-Blanco, A., Lucatello, S. 2006, \aap, 450, 523 

\bibitem[Carretta et al. (2009a)]{carretta09a}  Carretta, E. et al. 2009a, \aap, 505, 117  

\bibitem[Carretta et al. (2009b)]{carretta09b} Carretta, E., Bragaglia, A., Gratton, R.G., Lucatello, S. 2009b,  \aap, 505, 139  

\bibitem[Carretta et al. (2009c)]{carretta09c} Carretta, E., Bragaglia, A., Gratton, R.G., D'Orazi, V., Lucatello, S. 2009c, \aap, 508, 695  

\bibitem[Cohen(2004)]{cohen} 
Cohen, J.~G.\ 2004, \aj, 127, 1545

\bibitem[Cohen et al.(2002)]{cohen02} Cohen, J.~G., Briley, 
M.~M., \& Stetson, P.~B.\ 2002, \aj, 123, 2525 

\bibitem[D'Antona \& Caloi (2004)]{dantona04} D'Antona, F., Caloi, V. 2004, \apj,
611, 871

\bibitem[D'Antona \& Ventura(2007)]{dv07} D'Antona, F., \& Ventura, P.\ 2007, \mnras, 379, 1431 

\bibitem[D'Antona et al. (2002)]{dantona02} D'Antona, F.,  Caloi, V., Montalban, J., Ventura, P., Gratton, R.  2002, \aap, 395, 69

\bibitem[D'Antona et al. (2005)]{dantona05} D'Antona, F., Bellazzini, M., Caloi, V.,
Fusi Pecci, F., Galleti, S., Rood, R.T. 2005, \apj, 631, 868

\bibitem[Decressin et al. (2007)]{decressin07} Decressin, T., Meynet, G., Charbonnel C. Prantzos, N.,
Ekstrom, S. 2007, \aap, 464, 1029

\bibitem[Denisenkov \& Denisenkova (1989)]{den89} Denisenkov, P.A., Denisenkova, S.N. 1989, A.Tsir., 1538, 11

\bibitem[D'Orazi et al. (2010, in prep.)]{dorazi10} D'Orazi et al., in preparation

\bibitem[Gehren et al. (2004)]{gehren04} Gehren, T., Liang, Y.C., Shi, J.R., Zhang,
H.W., Zhao, G. 2004, \aap, 413, 1045

\bibitem[Goldoni et al. (2006)]{goldoni06} Goldoni P., Royer F., Franc¸ois P., Horrobin M., Blanc G., Vernet J.,
Modigliani A., Larsen J., 2006, in McLean I. S., IyeM., eds, Proc. SPIE
Vol. 6269, Ground-based and Airborne Instrumentation for Astronomy.
SPIE, Bellingham, p. 62692K

\bibitem[Gratton (1988)]{rosa} Gratton, R.G. 1988, Rome Obs. Preprint Ser., 29

\bibitem[Gratton et al. (1999)]{gratton99} Gratton, R.G., Carretta, E., Eriksson, K., \& Gustafsson, B.\ 1999, \aap, 350, 955 

\bibitem[Gratton et al. (2000)]{gratton00} Gratton, R.G., Sneden, C., Carretta, E., Bragaglia, A., 2000, A\&A, 354, 169

\bibitem[Gratton et al. (2001)]{gratton01} Gratton, R.G., Bonifacio, P., Bragaglia, A., et al. 2001, \aap, 369, 87

\bibitem[Gratton et al. (2004)]{gratton04} Gratton, R.G., Sneden, C., \& Carretta, E. 2004, \araa, 42, 385

\bibitem[Gratton et al. (2010)]{gratton10} Gratton, R.G., Carretta, E., Bragaglia,
A., Lucatello, S., D'Orazi, V. 2010,  \aap, in press arXiv:1004.3682

\bibitem[Harris (1996)]{harris96} Harris, W.~E. 1996, \aj, 112, 1487 

\bibitem[Ivans et al.(2001)]{ivans01} Ivans, I.~I., Kraft,  R.~P., Sneden, C., Smith, G.~H., Rich, R.~M., 
\& Shetrone, M.\ 2001, \aj, 122, 1438 

\bibitem[Langer et al. (1993)]{langer93} Langer, G.E., Hoffman, R., \& Sneden, C. 1993, \pasp, 105, 301

\bibitem[Lee (2010)]{lee10} Lee, J.-W.\ 2010, \mnras, 405, L36 

\bibitem[Moehler \& Sweigart(2006)]{moehler06} Moehler, S., \& Sweigart, A.~V.\ 2006, \aap, 455, 943 

\bibitem[Norris (2004)]{norris04} Norris, J.~E.\ 2004, \apjl, 
612, L25 

\bibitem[Pietrinferni et al.(2006)]{pietrinferni06}
Pietrinferni, A., Cassisi, S., Salaris, M., \& Castelli, F. \ 2006, \apj, 642, 797

\bibitem[Piotto (2009)]{piotto09} Piotto, G. 2009, ``The ages of stars", IAU Symp. 258, 233

\bibitem[Piotto et al. (2005)]{piotto05} Piotto, G., et al. 2005, \apj, 621, 777

\bibitem[Piotto et al. (2007)]{piotto07} Piotto, G., et al. 2007, \apjl, 661, L53

\bibitem[Ram{\'{\i}}rez \& Cohen(2003)]{rc03} Ram{\'{\i}}rez, S.~V., \& Cohen, J.~G.\ 2003, \aj, 125, 224 

\bibitem[Salaris \& Cassisi(2005)]{scbook} Salaris, M., \& Cassisi, S.\ 2005, 
Evolution of Stars and Stellar Populations, Wiley-VCH, Weinheim, Germany

\bibitem[Sbordone et al.(2005)]{sbordone} 
Sbordone, L., Bonifacio, P., Marconi, G., Buonanno, R., \& Zaggia, S.\ 2005, \aap, 437, 905 

\bibitem[Sneden(1973)]{moog}
Sneden, C. \ 1973, \apj, 184, 839

\bibitem[Vernet et al.(2009)]{vernet09} Vernet, J., D'Odorico, 
S., Christensen, L., Dekker, H., Mason, E., Modigliani, A., 
\& Moehler, S.\ 2009, The Messenger, 138, 4 

\bibitem[Villanova et al.(2009)]{villanova09} Villanova, S., Piotto, G., \& Gratton, R.~G.\ 2009, \aap, 499, 755 
\bibitem[Yong et al.(2009)]{yong09} Yong, D., Grundahl, F., 
D'Antona, F., Karakas, A.~I., Lattanzio, J.~C., 
\& Norris, J.~E.\ 2009, \apjl, 695, L62 

\end{thebibliography}
\end{document}